\documentclass[aps,pre,reprint,superscriptaddress, showpacs, showkeys]{revtex4-1}

\usepackage{amssymb,amsmath}

\usepackage[utf8]{inputenc}

\usepackage{epsfig}
\usepackage{bm}
\usepackage{color}
\usepackage{graphicx}
\usepackage{hyperref}
\usepackage{multirow}

\newcommand\beq{\begin{equation}}
\newcommand\eeq{\end{equation}}
\newcommand\beqa{\begin{eqnarray}}
\newcommand\eeqa{\end{eqnarray}}
\def\bal#1\eal{\begin{align}#1\end{align}}

\begin{document}
\title{Induced correlations and rupture of molecular chaos by anisotropic dissipative Janus hard disks}
\author{Antonio Lasanta}
\affiliation{Gregorio Mill\'an Institute of Fluid Dynamics, Nanoscience and Industrial Mathematics, 
Department of Materials Science and Engineering and Chemical Engineering, Universidad Carlos III de Madrid, Legan\'es, Spain}
\affiliation{Instituto de Energ\'{\i}as Renovables, Universidad Nacional Aut\'onoma de M\'exico (U.N.A.M.), Temixco, Morelos 62580, M\'exico}
\author{Aurora Torrente}
\affiliation{Gregorio Mill\'an Institute of Fluid Dynamics, Nanoscience and Industrial Mathematics, Department of Materials Science and Engineering and Chemical Engineering, Universidad Carlos III de Madrid, Legan\'es, Spain}
\author{Mariano L\'{o}pez de Haro}
\affiliation{Instituto de Energ\'{\i}as Renovables, Universidad Nacional Aut\'onoma de M\'exico (U.N.A.M.), Temixco, Morelos 62580, M\'exico}

\date{\today}

\begin{abstract}
A system of smooth ``frozen" Janus-type disks is studied. Such disks cannot rotate and are divided by their diameter into two sides of different inelasticities. Taking as a reference a system of colored elastic disks, we find differences in the behavior of the collisions once the anisotropy is included. A homogeneous state, akin to the homogeneous cooling state of granular gases,  is seen to arise and the singular behavior of both the collisions and the precollisional correlations are highlighted.
\end{abstract}



\maketitle


\section{Introduction} \label{intro}

Janus particles are characterized by presenting two or more different properties on their surfaces. Due to the fact that such properties allow them to have different composition and functionality \cite{PG17}, Janus particles offer a variety of potential applications such as drug design, biomarkers, bactericides and many others \cite{LL17}. During the past few years, theoretical and experimental studies on Janus particles, as well as on their synthesis, have represented a major challenge for the scientific community (see, for instance, Refs.\ \cite{GMMN13,RMDF14,GS14,ZLG15,GBB16,MVC18}). Most studies have focused their attention on equilibrium systems and the study of their phase behavior \cite{F13}. On the other hand, from a nonequilibrium perspective, only colloidal particles interacting in a solvent have been considered, thus accounting for hydrodynamic interactions but neglecting the particles' inertia \cite{GMMN13,GS14,ZLG15,GBB16,MVC18}.

It is interesting to point out that, with the proper adjustment of the parameters, a system of Janus particles embodies analogies with the so-called `microswimmers', which are the prototype systems of active matter \cite{F13, ANVP19}. These swimmers, which constitute presently a `hot' and highly attractive research topic within statistical physics, are capable of self-propelling without any external energy input. Janus particles may also be employed to investigate melting and appearance of nematic phases in disks with simple inhomogeneous properties as is the case of the discotic liquid crystals \cite{WWKKKLHSBGL15}. The previous background serves as a motivation for the present study. Our aim is to consider a dilute or moderately dilute fluid of 2 dimensional Janus particles (disks) in which these collide obeying a collision rule that reflects their asymmetry, namely, a rule in which nonhomogeneous restitution coefficients are incorporated.

Among the tools that are available to study the physical properties of fluid systems, computational simulation techniques occupy a prominent place \cite{R04,B94}. Perhaps one of the most popular ones is molecular dynamics (MD) and, in the case of the less dense systems, Monte Carlo (MC) simulations of either the corresponding Boltzmann or Enskog equation. Such techniques have proven to be useful not only in the analysis of a variety of transport problems in classical fluids \cite{R04,B94,GS03} but also in the successful description of the behavior of complex fluid systems and granular media \cite{R04,PS05,P14}. In the context of this paper, a granular medium should be understood as a system composed of a huge number of particles of mesoscopic size that have mutual inelastic collisions. That is, the particles loose kinetic energy when they collide \cite{G03,AT06}. This loss of kinetic energy is in turn absorbed by the internal degrees of freedom of the particles of the material medium, such degrees of freedom being totally uncoupled to the dynamics of the mesoscopic particles. Given the intrinsic characteristics of these systems, a convenient way to study their physical behavior is through MD simulations, in which the (event-driven) algorithms \cite{AT89} are adjusted to account for the fact that the collisions are inelastic. With such an approach, it has been possible to correctly describe laminar flow problems, instabilities, phase transitions, statistical correlations, diffusion, segregation and other phenomena occurring in granular media \cite{G03,AT06}. Furthermore, in the case of low and moderate densities, a number of papers have demonstrated that the single-particle time-dependent velocity distribution function obeys either the Boltzmann or Enskog kinetic equations \cite{G03}. The fundamental ingredient behind both  equations is the stosszahlansatz or molecular chaos assumption. It should be stressed that kinetic theory has proven to be a very accurate and powerful tool to investigate dilute granular gases made of isotropic particles \cite{G19,P15}. Moreover, the use of the direct MC simulation (DSMC) method has also been successfully applied in the case of granular media \cite{G03,AT06,MS00,MG02}. In fact, two of us have contributed \cite{LVPS17,VLSG17,TLLVPS19} to the simulation, using either MD or DSMC, of some transport problems in dilute granular gases, including the segregation of granular intruders immersed in a granular gas of rough spheres \cite{VLSG17}, a system which has received substantial recent attention in the literature \cite{BPKZ07,VS15}. The aim of this paper is to profit from the analogies between granular media and media composed by anisotropic Janus-like particles and the computational techniques that have been employed to describe the physical properties of the former to study a dilute gas of Janus-like granular disks. 

The paper is organised as follows. In order to have a proper perspective, in Section~\ref{s2} we provide a brief description of the main results stemming out of the kinetic theory of a dilute gas of inelastic isotropic granular disks, including the so-called homogeneous cooling state (HCS). In Section~\ref{s3} we characterize our system and the collision rule for Janus-like granular disks, where we consider varying values of the coefficients of restitution corresponding to both sides of the 2D particles and, in Section~\ref{s4}, we report our findings from numerical experiments, involving velocity distribution functions, energy time evolution, cooling rates and  kurtosis.  We close the paper in Section~\ref{s5} with some concluding remarks.


\section{The homogeneous cooling state of a dilute gas of isotropic granular disks}
\label{s2}

We begin by considering a system of $N$ smooth inelastic hard disks,
 of mass $m=1$ and  diameter $\sigma=1$. 
The interaction between disks is characterized by the rule:  
\begin{eqnarray}
{\bf v}_{i}^{*}&=&{\bf v}_{i}-\frac{1+\alpha}{2} ({\bf g}\cdot
\widehat{\mbox{\boldmath$\sigma$}}) \widehat{\mbox{\boldmath$\sigma$}}
\, , \nonumber \\
{\bf v}_{j}^{*}&=&{\bf v}_{j}+\frac{1+\alpha}{2} ({\bf g}\cdot
\widehat{\mbox{\boldmath$\sigma$}})\widehat{\mbox{\boldmath$\sigma$}} \, ,
\end{eqnarray}
where we have introduced the constant coefficient of normal  restitution $\alpha$,  the asterisks denoting velocities after the collision, ${\bf g}={\bf v}_{i}-{\bf v}_{j}$ being the relative velocity, and $\widehat{\mbox{\boldmath$\sigma$}}$ a unit vector joining the centers of particles $i$ and $j$ at contact.

The appearance of the restitution coefficient $\alpha$ accounts for the loss of energy of particles $i$, $j$ after the collision. Note that $\alpha \leq 1$, and that the value $\alpha=1$  would correspond to elastic collisions. 

The equation that properly describes this system is the Boltzmann equation of a system of freely evolving hard disks, which reads \cite{Brey96,Brey04}
\begin{equation}
\left( \frac{\partial}{\partial t}+{\bf v}\cdot{\boldmath \nabla}\right)
f({\bf r},{\bf v},t)=
{\cal C}_{B}[{\bf r},{\bf v}|f({\bf r},{\bf v},t) ] \, ,
\end{equation}
with $f({\bf r},{\bf v},t)$ denoting the single-disk velocity distribution function and ${\cal C}_{B}$ the inelastic Boltzmann collision operator, namely
\begin{equation}
\begin{aligned}
{\cal C}_{B}[{\bf r},{\bf v}|f({\bf r},{\bf v},t) ]=&\\
\int d{\bf v}_{1} \int d\widehat{\mbox{\boldmath$\sigma$}}\,
\theta ({\bf g}\cdot \widehat{\mbox{\boldmath$\sigma$}})
({\bf g}\cdot \widehat{\mbox{\boldmath$\sigma$}})&
(\alpha^{-2} b^{-1}-1)f({\bf r},{\bf v},t) f({\bf r},{\bf v}_{1},t)\, ,\\
\end{aligned}
\end{equation}
$\theta$ is the Heaviside step function and $b^{-1}$ an operator transforming velocities ${\bf v}$ and  ${\bf v}_{1}$ to its right into their precollisional values. As it occurs in the purely elastic case, the derivation of this equation is based on the molecular chaos hypothesis. This means that the two-disks distribution function may be factorised in the precollisional disks as
$f^{(2)} ({\bf x}_{1},{\bf x}_{2},t)= f ({\bf x}_{1},t) f ({\bf x}_{2},t)$,
where ${\bf x}_{i}\equiv\{ {\bf r}_{i},{\bf v}_{i} \}$. 

It is well known that when a system of inelastic particles such as the one described above is allowed to evolve freely, it reaches a homogeneous state with no fluxes or gradients, namely, the homogeneous cooling state.  Such state is characterised by a single macroscopic variable, specifically the temperature $T(t)$, defined as proportional to the average kinetic  energy. This variable evolves according to the so called Haff's law \cite{H83},
\begin{equation}
T(t)=\dfrac{T(0)}{\left(1+\dfrac{t}{t_{0}}\right)^{2}} \, 
\end{equation}
where $t_{0}$ is the time characterizing the energy decay.

In the HCS, the Boltzmann equation admits a solution  $f({\bf v},t)$ that  obeys the scaling law \cite{Brey96,Brey04,Re77}
\begin{equation}
f({\bf v},t)=\frac{n_{H}}{v_{0}(t)^{2}} \ \varpi \left(\frac{{\bf v}}{v_{0}(t)}
\right) \, ,
\end{equation}
where $n_{H}$ is the homogeneous density, and $v_{0}=\sqrt{2 k_{B} T}$ is the thermal velocity of the system, $k_{B}$ being the Boltzmann constant. Note that for  the time evolution of the system, the relevant hydrodynamic field is the temperature, which corresponds to the second moment of the velocity, namely $\langle v^2 \rangle$. On the other hand the function $\varpi$, which follows from the Boltzmann equation, is given by

\begin{equation}
\varpi({\bf c})=\pi e^{-c^{2}} \sum_{j\geq 0} a_{j} 
S^{(j)} (c^{2}) \, , \mbox{\hspace{10mm}}  {\bf c}=\frac{{\bf v}}{v_{0}} \, 
\end{equation}
where $S^{(j)}$ are  the Sonine polynomials, whose expressions can be found in \cite{Re77}.

In our case the relevant coefficient is $a_{2}$, which is related to the kurtosis of the velocity distribution function, that is
\begin{equation}
a_{2}=\frac{1}{2} \left[ \frac{\langle v^{4}\rangle}{\langle v^{2}
\rangle^{2}} -2 \right] \, .
\label{eq:a2}
\end{equation}
It should be stressed that $a_{2}$ is very small, which means that the function $\varpi$ is very close to the Gaussian distribution \cite{Brey96}. The non-Gaussianity is reflected in the exponential tails.

The expression for $t_{0}$ may be obtained from the Boltzmann equation \cite{BDKS98}, and reads:
\begin{equation}
t_{0}^{-1}\simeq \frac{1}{2}(1-\alpha^{2})
\left( k_{B} \pi
T(0)\right)^{1/2} n_{H} \, .
\end{equation}

To close this section, and for later purposes, it is convenient to express Haff's law  in terms of the average number of collisions per particle, $\tau$, that is:
\begin{equation}
T(\tau)=T(0) e^{-(1-\alpha^{2}) \tau} \, ,
\label{eq:Haff}
\end{equation}
which, interestingly, states that the kinetic energy decays exponentially with the number of collisions.


\section{The anisotropic Janus-like disks case}
\label{s3}

\begin{figure}[!b]
  \centering
  \includegraphics[angle=0,width=8.6cm]{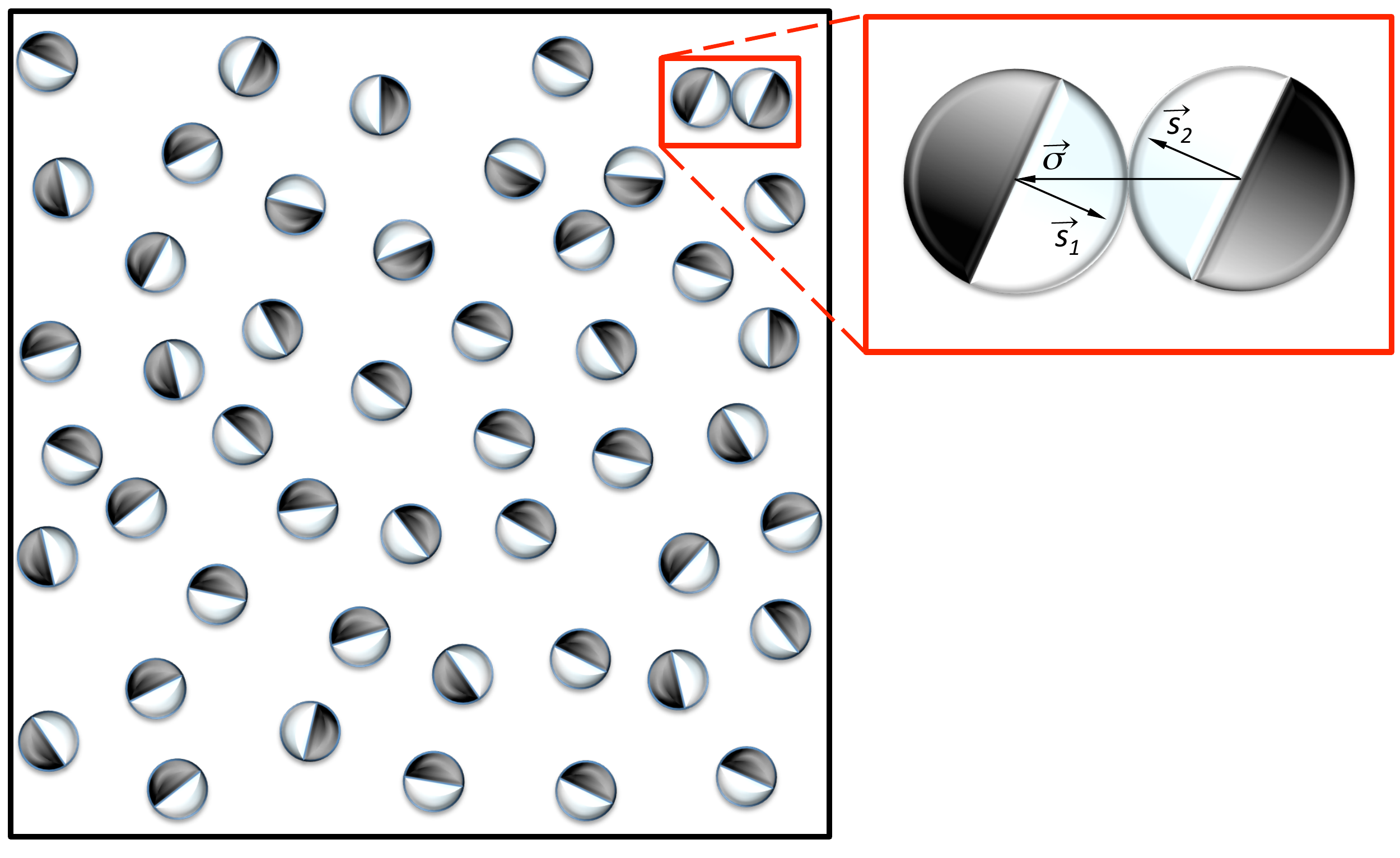}
 \caption{Schematic representation of the collision between two Janus particles in a dilute fluid; different halves are color-coded as ``black'' and ``white''. Vectors ${\bf s}_{1}$ and ${\bf s}_{2}$ are perpendicular to the diameter defining both sides.}
 \label{esquema}
 \end{figure}

Let us consider now a set of $N$ smooth anisotropic disks. Each disk comprises two parts of equal size on each side of its diameter, which are characterized by a value of the coefficient of restitution. Such values may be the same or different for both sides and either equal to one, less than one or greater than one. The disks may collide with one another in three different forms depending on which side of each disk takes part in the collision. A schematic representation of the collision is depicted in Fig.\ \ref{esquema} and the applicable collision rule is the following

\begin{eqnarray}\label{collisionRule}
  &{\bf v}'_{1}={\bf v}_{1}+\dfrac{1+\alpha_{1} \left( {\bf s}_{1} , {\bf s}_{2} \right)}{2}
 \widehat{\mbox{\boldmath$\sigma$}} \left( {\bf g}\cdot \widehat{\mbox{\boldmath$\sigma$}} \right) , \nonumber\\
 &{\bf v}'_{2}={\bf v}_{2}+\dfrac{1+\alpha_{2}\left({\bf s}_{1},{\bf s}_{2} \right)}{2} \widehat{\mbox{\boldmath$\sigma$}}\left(  {\bf g}\cdot
\widehat{\mbox{\boldmath$\sigma$}} \right).
 \end{eqnarray}
Here, ${\bf v}'_{i}$ indicates postcollisional velocities, ${\bf g}={\bf v}_{1}-{\bf v}_{2}$,
 $\widehat{\mbox{\boldmath$\sigma$}}=\dfrac{{\bf r}_{1}-{\bf r}_{2}}{\left| {\bf r}_{1}-{\bf r}_{2} \right|}$ and $\alpha_1 \left({\bf s}_{1},{\bf s}_{2} \right)$ and $\alpha_2 \left({\bf s}_{1},{\bf s}_{2} \right)$ are the coefficients of restitution which, as stated above, depend on the orientation of the particles at the time of the collision as given by the vectors ${\bf s}_{1}$ and ${\bf s}_{2}$. Such vectors are assigned to each of the $N$ particles in a direction randomly distributed and perpendicular to the diameter delimiting the two (different) sides. 

The dependence on $\alpha_i  \left({\bf s}_{1},{\bf s}_{2} \right)$ may be parameterized by labeling with colors, $W$ and $B$, the ``white" and ``black" sides, respectively, as follows
 \begin{eqnarray}
  &{\bf s}_{1}\cdot \widehat{\mbox{\boldmath$\sigma$}} \ge   0 \rightarrow  \text{disk 1 side }W \\
  & {\bf s}_{1}\cdot \widehat{\mbox{\boldmath$\sigma$}} <   0 \rightarrow  \text{disk 1 side }B \\
  & {\bf s}_{2}\cdot \widehat{\mbox{\boldmath$\sigma$}} <   0 \rightarrow  \text{disk 2 side }W  \\
  &{\bf s}_{2}\cdot \widehat{\mbox{\boldmath$\sigma$}} \ge  0 \rightarrow    \text{disk 2 side }B
  \end{eqnarray}

An ``event-driven'' molecular dynamics algorithm has been developed to analyze the time evolution of such a system. We have taken a square box with $N$ disks, whose diameter is $\sigma$, and a given number density $\rho$, which determines the size of the box. We have considered $N=k^2$ for some $k\in\mathbb{N}$ and divided the box into $N$ square cells. Initially, the disks are centrally placed in the cells and are assigned a random velocity that follows a Maxwell-Boltzmann distribution;  then they are made to elastically collide with each other until an equilibrium state is reached. After such a transient period, an anisotropy vector ${\bf s}_i$, whose direction is distributed uniformly in the interval $\left[0,2\pi \right[$, is assigned to each disk. Once this has been done, the disks start to collide following the collision rules (\ref{collisionRule}). Of course, the coefficients of restitution $\alpha_{BB}$, $\alpha_{WW}$, $\alpha_{BW}$ and $\alpha_{WB}$, where the subindices state the sides at contact during the collision, may be freely chosen. In particular, we have defined $\alpha_{WB}=\alpha_{BW}= \dfrac{\alpha_{BB}+\alpha_{WW}}{2}$.

The outcome of our simulations comprises the measurement of the velocity distribution functions for the particles, the time evolution of the energy and the kurtosis. Furthermore, we are also able to evaluate the fraction of each type of collision (and hence measure the precollisional correlations), the time evolution of the different temperatures present in the system and the cooling rates for each type of collision.


\section{Results}\label{s4}

\begin{figure}[!h]
  \centering
  \includegraphics[angle=0,width=8.6cm]{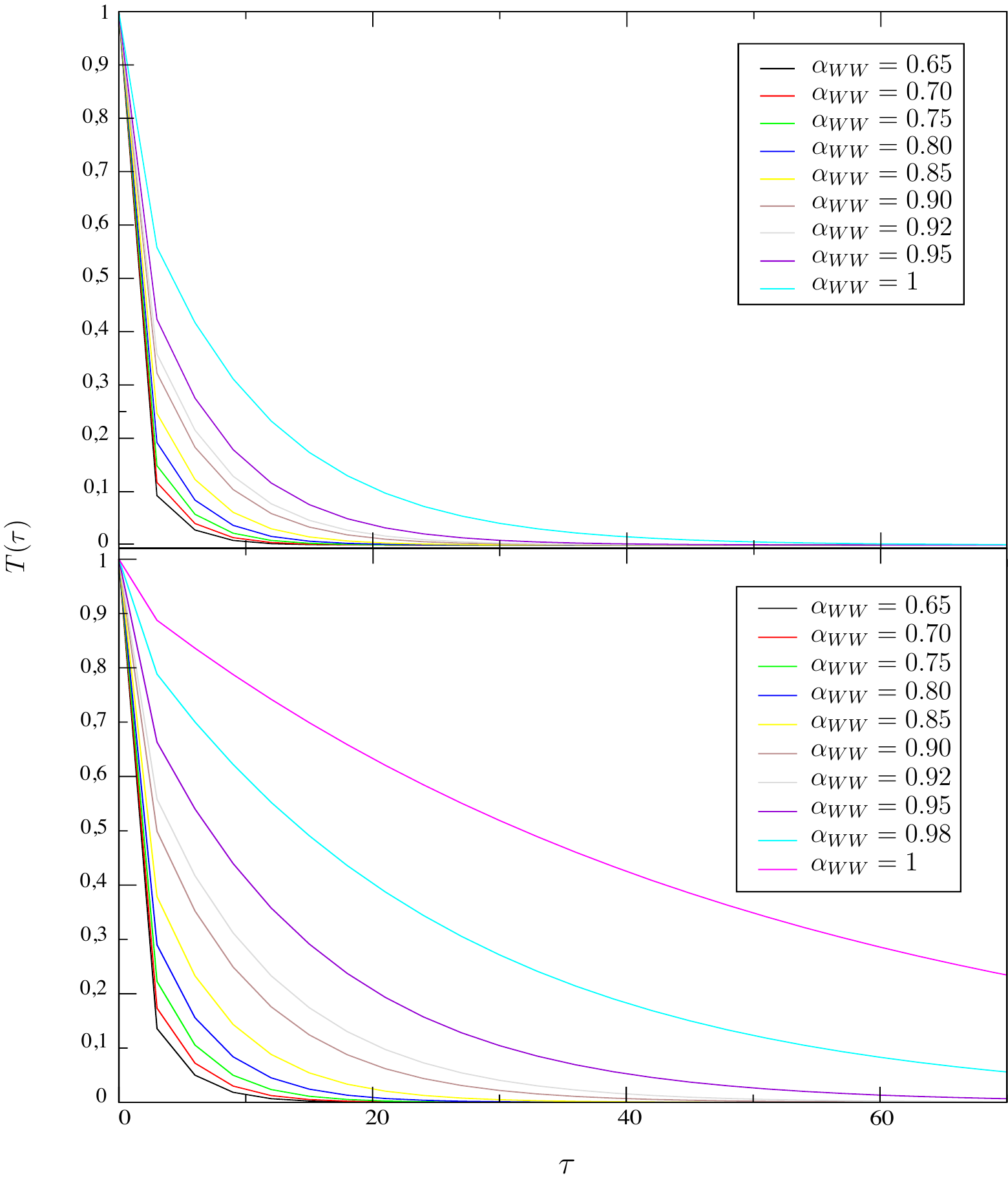}
 \caption{Temperature vs number of collisions per particle, $\tau$, for $\alpha_{BB}=0.9$ (top panel) and  $\alpha_{BB}=0.98$ (bottom panel) for different values of $\alpha_{WW}$ in the interval $[0.6, 1.0]$. }
 \label{energia}
 \end{figure}

For the sake of having some reference values for later comparison with the collisional dynamics of our system of interest, and since, to the best of our knowledge, there are no previous studies in the literature of elastic colored disks, we begin with the fully elastic case. In this instance we have considered a system of 1225 disks with a number density of 0.01 and taken an average over 180 trajectories. We have taken systems with a fixed value of the coefficient of restitution $\alpha_{BB}$ (namely, $\alpha_{BB}=0.9$  and $\alpha_{BB}=0.98$) and for such systems then we have varied the value of $\alpha_{WW}$ in the range $\left[0.65,1.0\right]$. In Fig.~\ref{energia} we display the time evolution of the temperature, time being measured as collisions per particle. In this figure, averages over the particles and trajectories have been performed. The system obeys Haff's law, as seen from the exponential decay.

\begin{figure}
  \centering
  \includegraphics[angle=0,width=8.6cm]{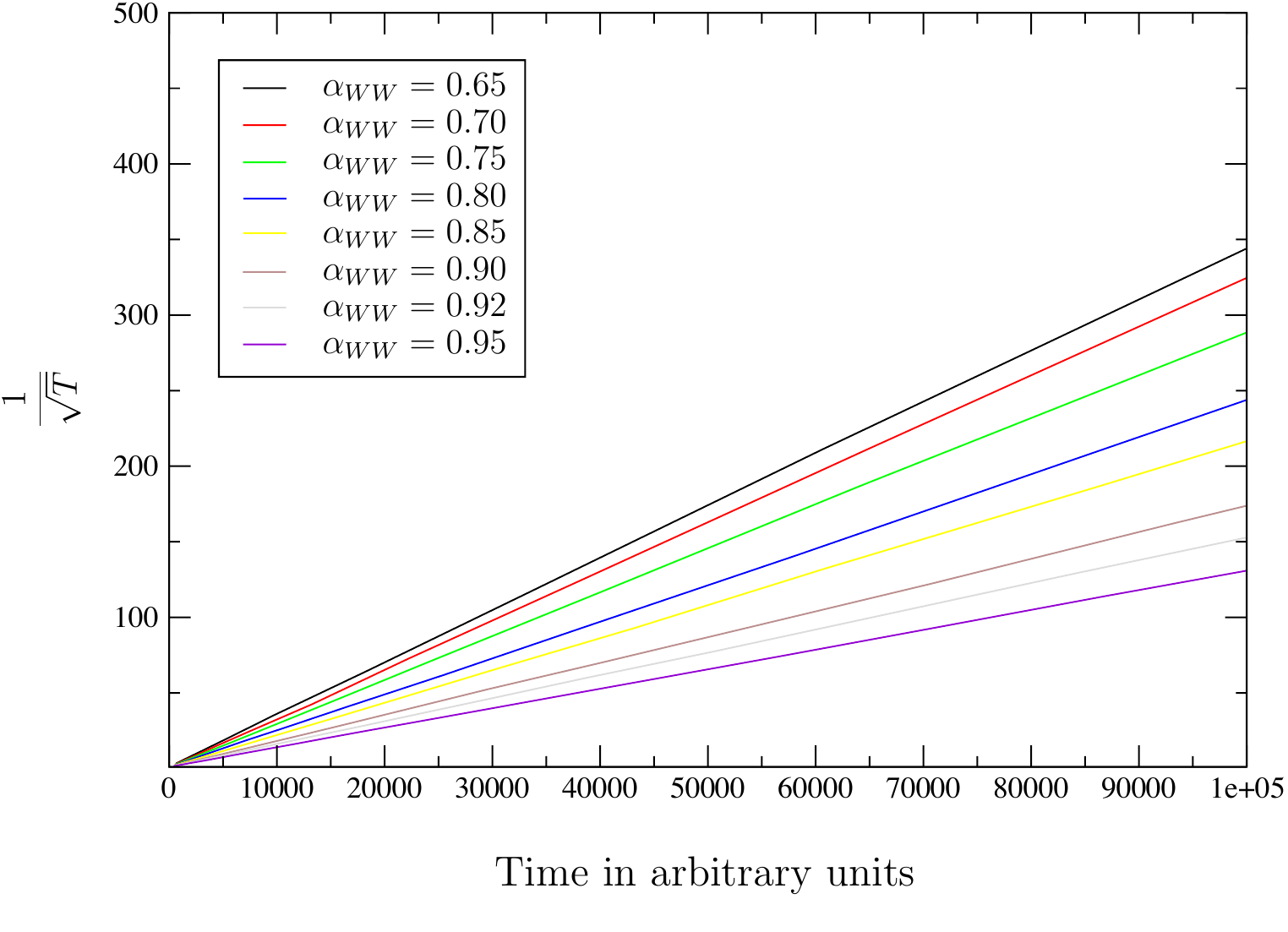}
  \caption{Haff's law for systems with  $\alpha_{BB}=0.90$ and different values of  $\alpha_{WW}$. The slopes of the straight lines decrease as $\alpha_{WW}$ increases.}
 \label{sqrtenergia}
  \end{figure}

\begin{figure}[!h]
  \centering
   \includegraphics[angle=0,width=8.6cm]{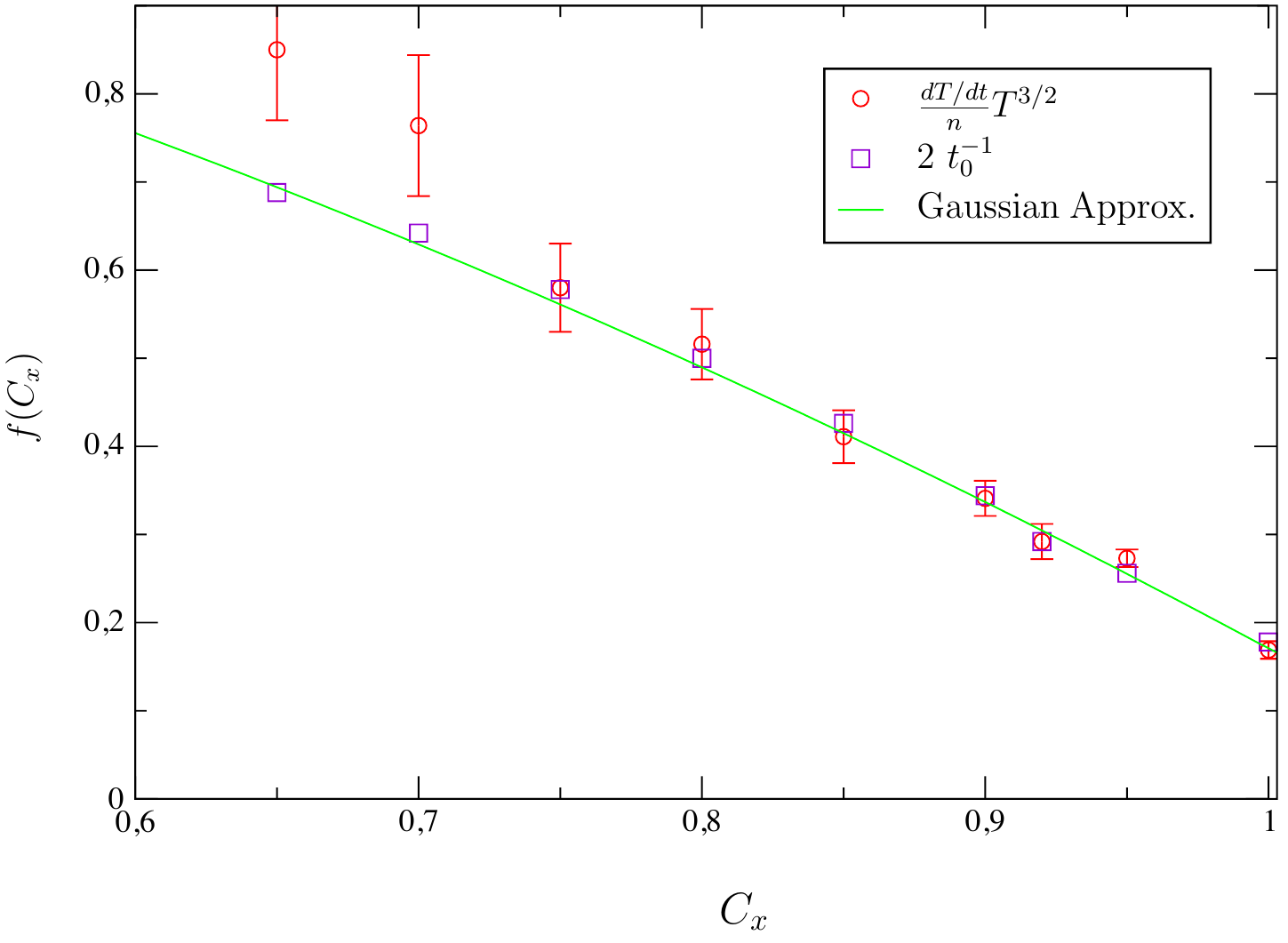}
 \caption{Values of the total cooling rates for $\alpha_{BB}=0.9$ and
 $0.65\le \alpha_{WW} \le 1.0$ (open red circles), along with their confidence limits,  compared with the Gaussian approximation (continuous line) as well as with the slopes of the straight lines in Fig.\ \ref{sqrtenergia} (open magenta squares).    
{\color{red} }
}
 \label{coolin2}
 \end{figure}

This fact is more evident in Fig.\ \ref{sqrtenergia}, where straight lines of the form $T\left(t\right)=T\left(0\right)\left(1+\frac{t}{t_0}\right)^{-2}$ are obtained for $\alpha_{BB}=0.90$ and different values of $\alpha_{WW}$.  In Fig.~\ref{coolin2} we present the values of $t_{0}^{-1}$ derived from the fit of the lines in Fig.~\ref{sqrtenergia} together with the cooling rates that have been measured directly and the Gaussian approximation for the value of  $t_{0}^{-1}$.

\begin{figure}[h]
  \centering
  \includegraphics[angle=0,width=8.6cm]{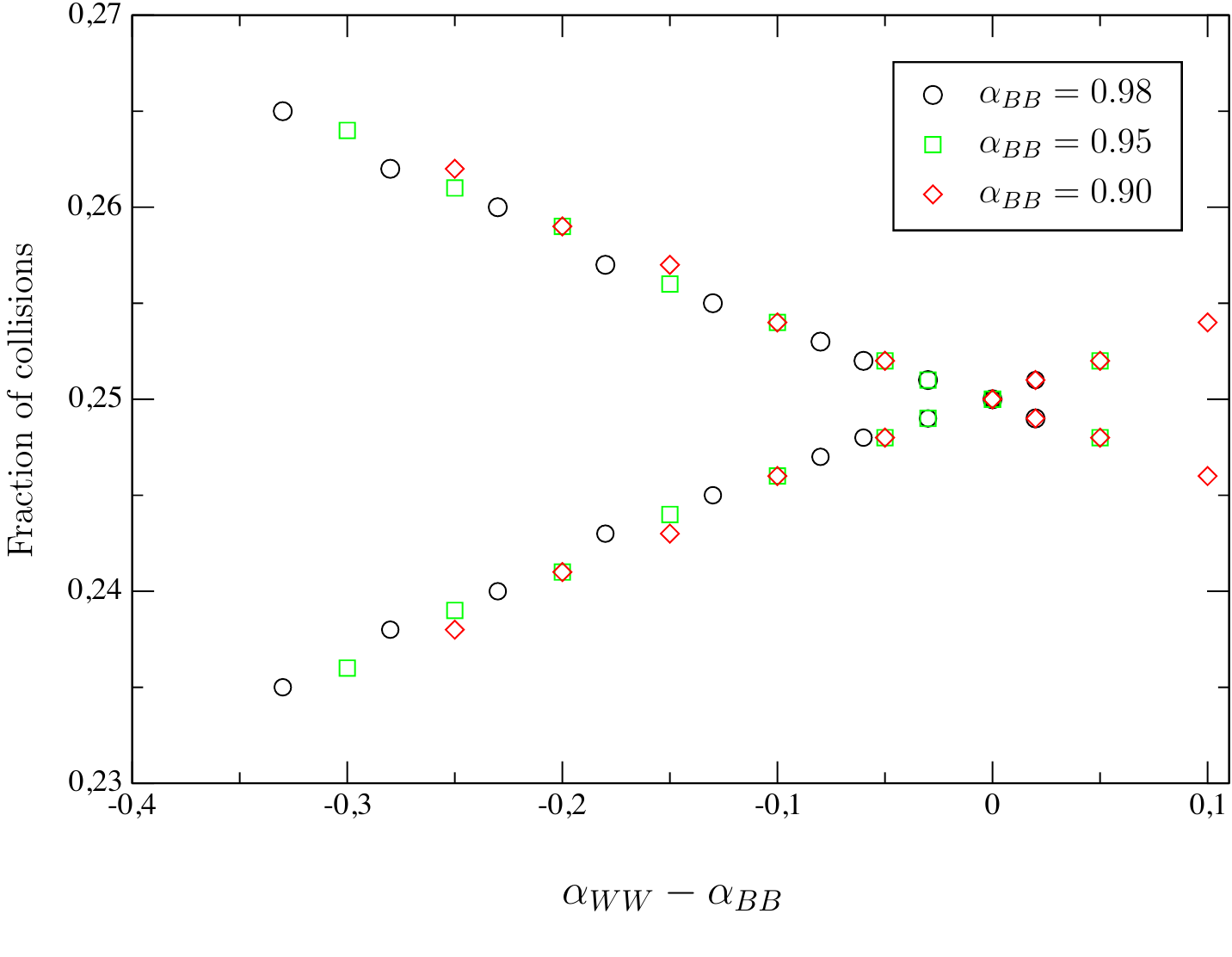} 
 \caption{Values of the fraction of collisions of each type vs the difference  $\alpha_{WW}-\alpha_{BB}$. Note the collapse into a single curve for varying values of $\alpha_{BB}$. The curve with increasing slope corresponds to that of $WW$ type.}
 \label{fracccolldif}
 \end{figure}

A remarkable feature present in our system is the fact that, when it comes to collisions of sides of the same color, there are more collisions of one color than of the other. In order to quantify this effect, we have measured the fraction of each type of single-color collision with respect to the total number of collisions that occur in the system  all along its evolution. We find that the more inelastic one color is the more collisions of this type take place. Further, and somewhat surprisingly, this number depends only on the difference between the elasticities of the sides, as observed in Fig.~\ref{fracccolldif}. This leads us to suggest that different temperatures of the disks in our system may arise depending on the color of their previous collision. Also, one can see that once a disk has undergone  a collision with one of its colored sides, the probability that it will collide with the opposite side (different color) is greater than the one that it will do it with the same side. This is due to the fact that such probability of collision is proportional to the relative velocities and these vectors are directed with more probability to the side opposite to the one the collision took place with.

\begin{figure}
 \centering
 \includegraphics[angle=0,width=8.6cm]{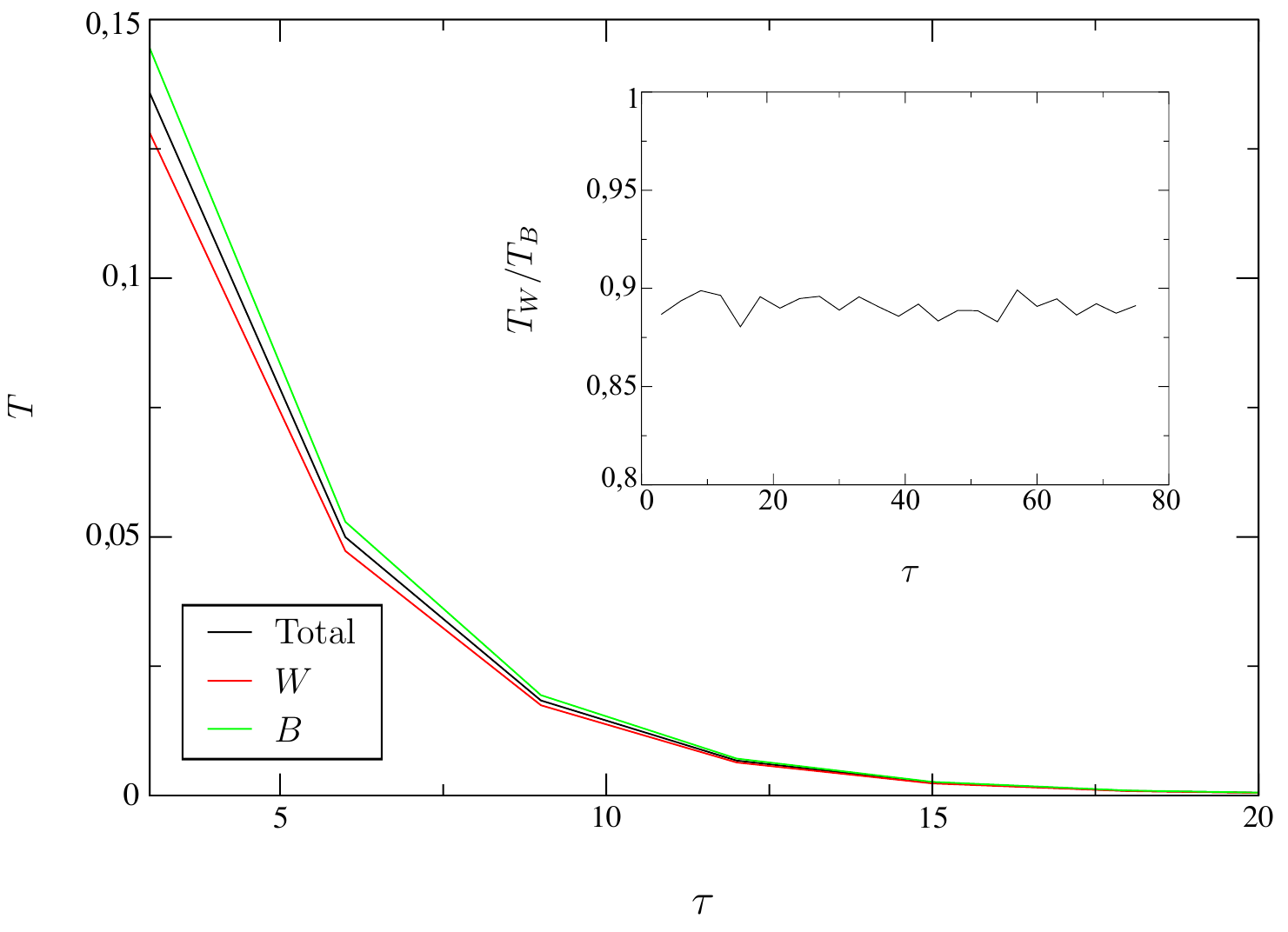}
 \caption{Values of the temperatures of each type vs the number of collisions per particle and in a range of these between 0 and 20 for an easy visualization of the difference, for $\alpha_{WW}=0.65$ and $\alpha_{BB}=0.98$. Note the appearance of three ``temperatures" in this case. In the inset, corresponding value of the ratio $T_W/T_B$ between the temperatures of each kind as a function of the number of collisions per particle. As pointed out in the text, this value is a constant.}
 \label{temtip098065}
 \end{figure}

\begin{figure}[!h]
 \centering
 \includegraphics[angle=0,width=8.6cm]{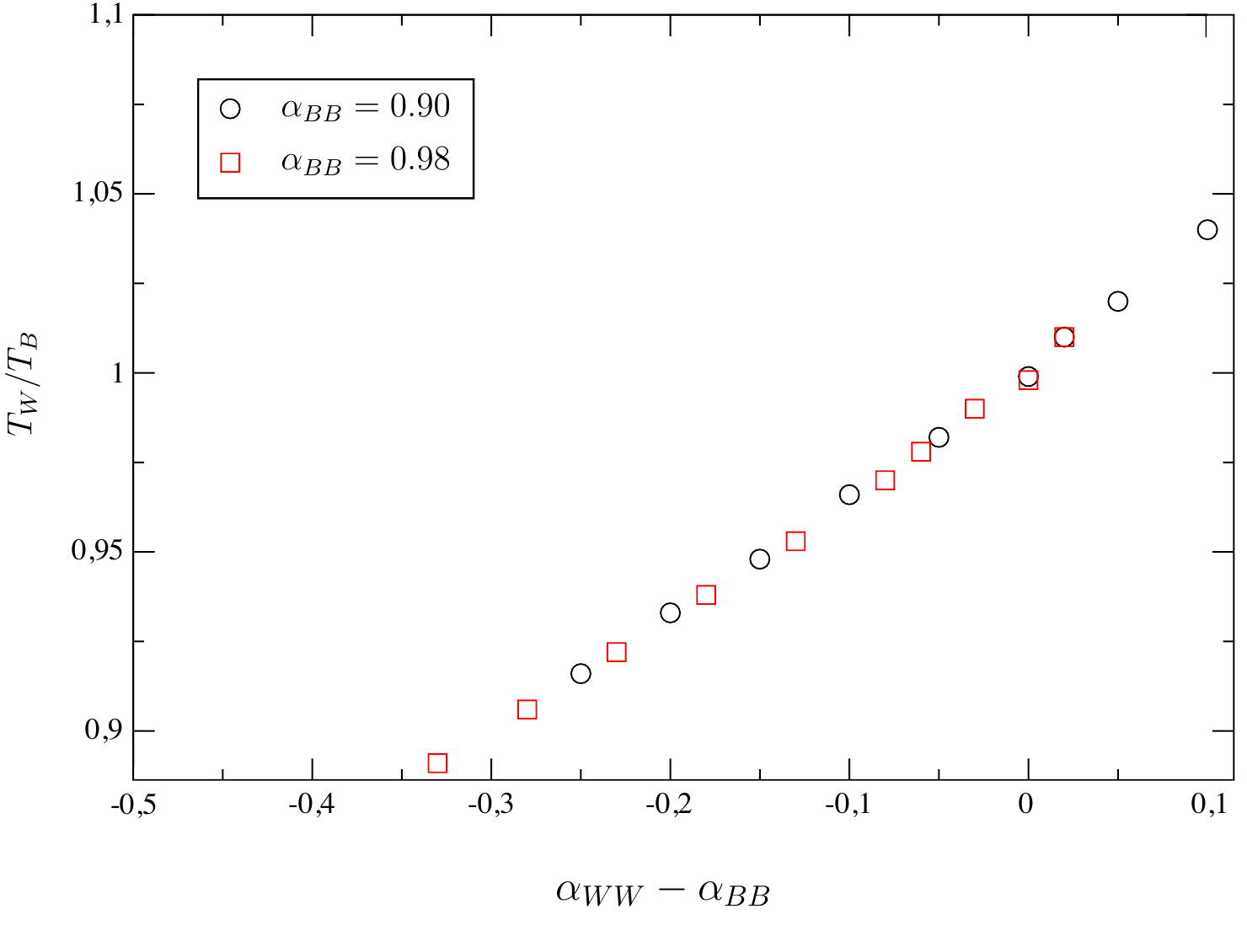}
 \caption{Value of the ratio between the temperatures of each kind as a function of the difference $\alpha_{WW}-\alpha_{BB}$. Note again the collapse of the curves.}
 \label{cocientet}
 \end{figure}

If we measure time by collisions per particle, at any temporal point we can consider the particles that have collided either with one side (color) or with the other and investigate two aditional ``temperatures", $T_W$ and $T_B$, each one corresponding to one color. In Fig.~\ref{temtip098065} it can be immediately seen that the particles that have collided with the black side have a greater temperature than those that have collided with the white side. This is not surprising since the white side has a coefficient of restitution $\alpha_W < \alpha_B$.  On the other hand, we also find that the ratio between both temperatures $T_W$ and $T_B$ is a constant (\emph{cf.} inset in Fig.~\ref{temtip098065}) and that it depends only on the difference $\alpha_{WW}-\alpha_{BB}$ (\emph{cf.} Fig.~\ref{cocientet}).

\begin{figure}[!h]
  \centering
  \includegraphics[angle=0,width=8.6cm]{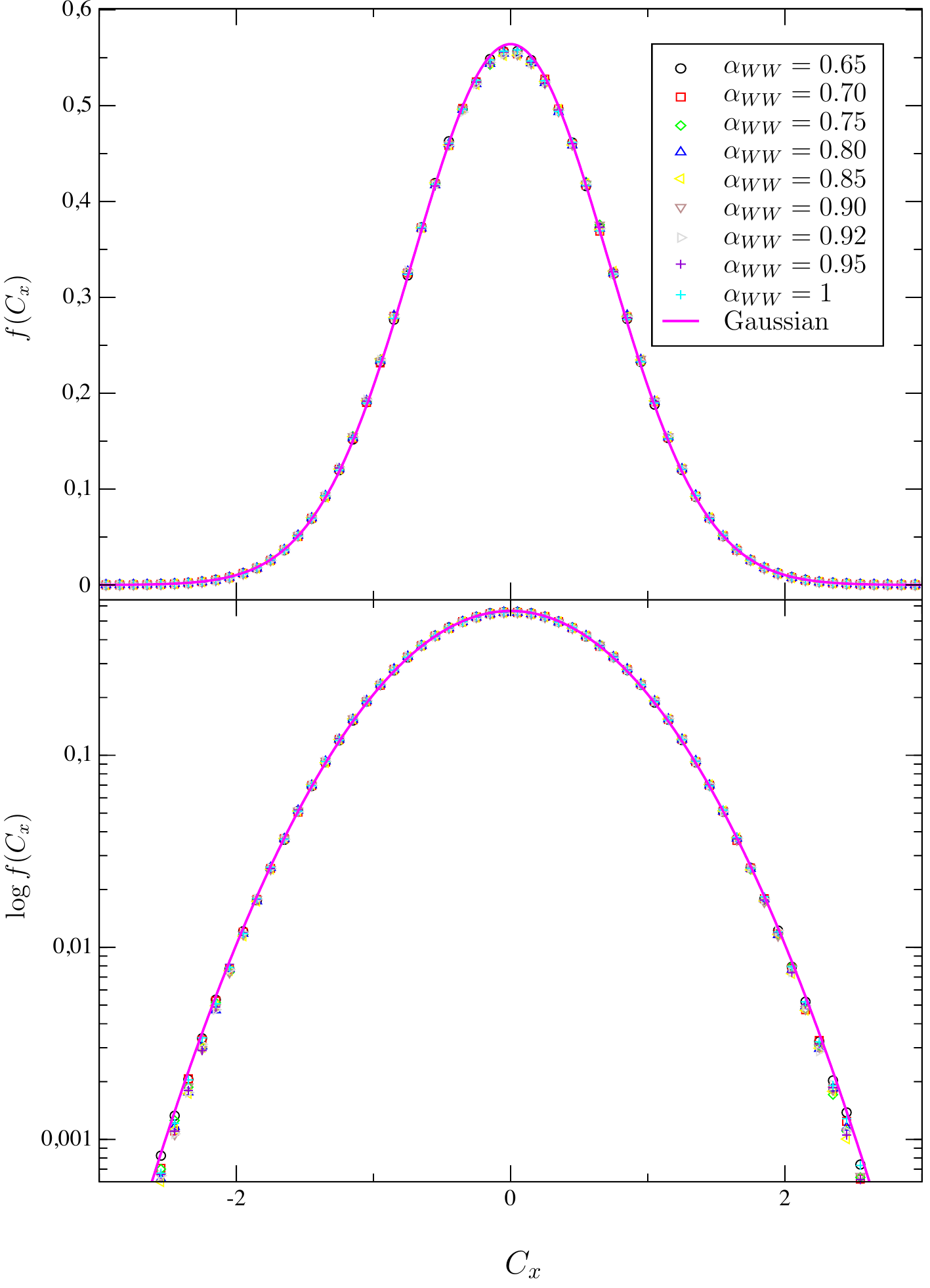}
 \caption{Velocity distribution functions (top panel) and their logarithm (bottom panel) for $\alpha_{WW}=0.9$ and $0.65 \le \alpha_{BB} \le 1.0$. The continuous line is the Maxwell-Boltzmann distribution.}
 \label{funciondistri}
 \end{figure}

\begin{figure}[!h]
  \centering
  \includegraphics[angle=0,width=8.6cm]{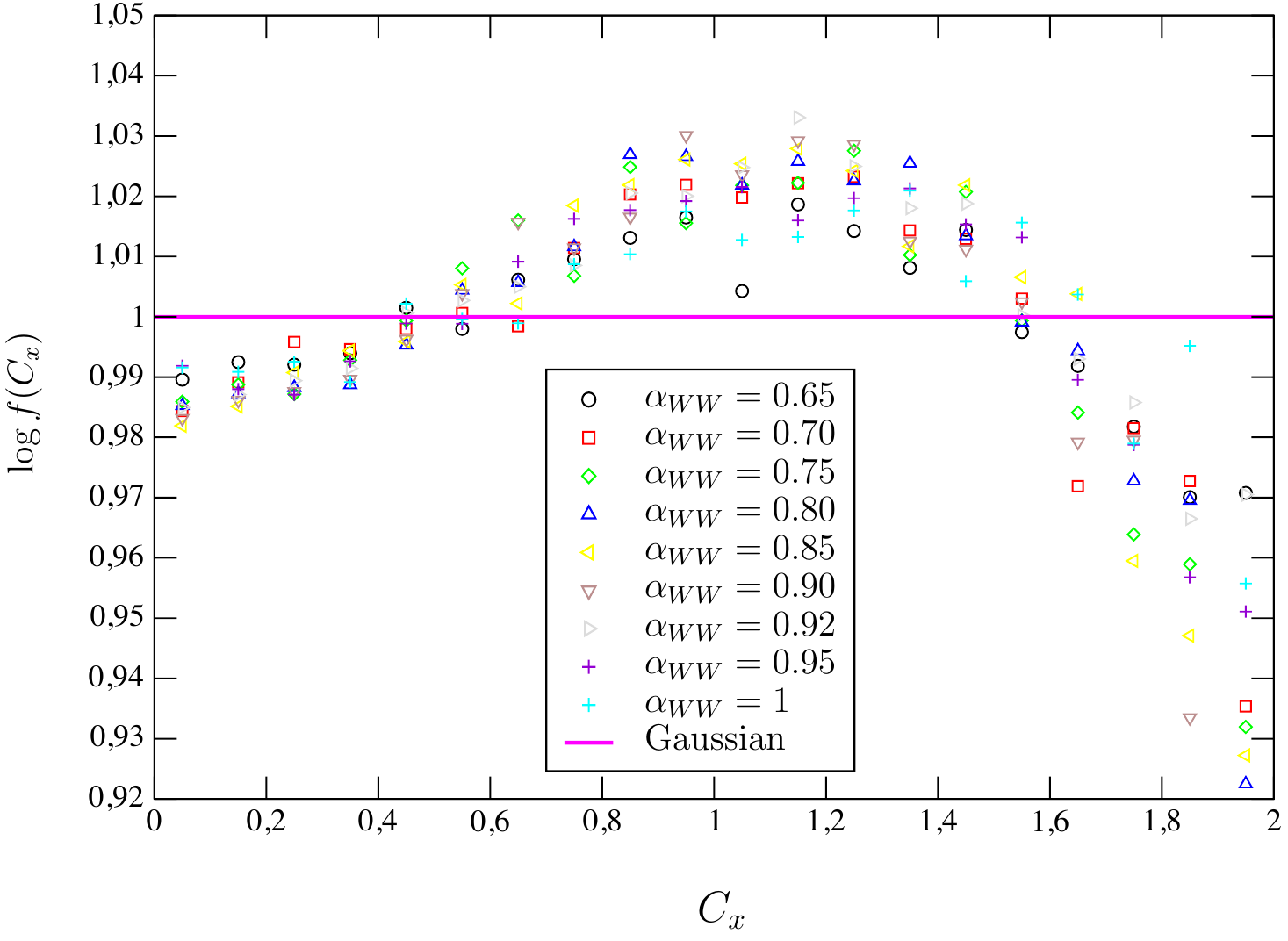}
 \caption{Tails of the velocity distribution functions divided by the Gaussian distribution for $\alpha_{WW}=0.9$ and $0.65 \le \alpha_{BB} \le 1.0$.}
 \label{funciondistribgauss}
 \end{figure}

 The velocity distribution functions corresponding to $\alpha_{BB}=0.9$ are shown in Fig.~\ref{funciondistri}. To a good approximation, they may be taken as Gaussian distributions. Also, the tails of the scaled distributions are displayed in Fig.~\ref{funciondistribgauss}. All these curves are similar to the ones obtained in isotropic granular media \cite{Brey96, Brey04}.

\begin{figure}[!h]
  \centering
  \includegraphics[angle=0,width=8.6cm]{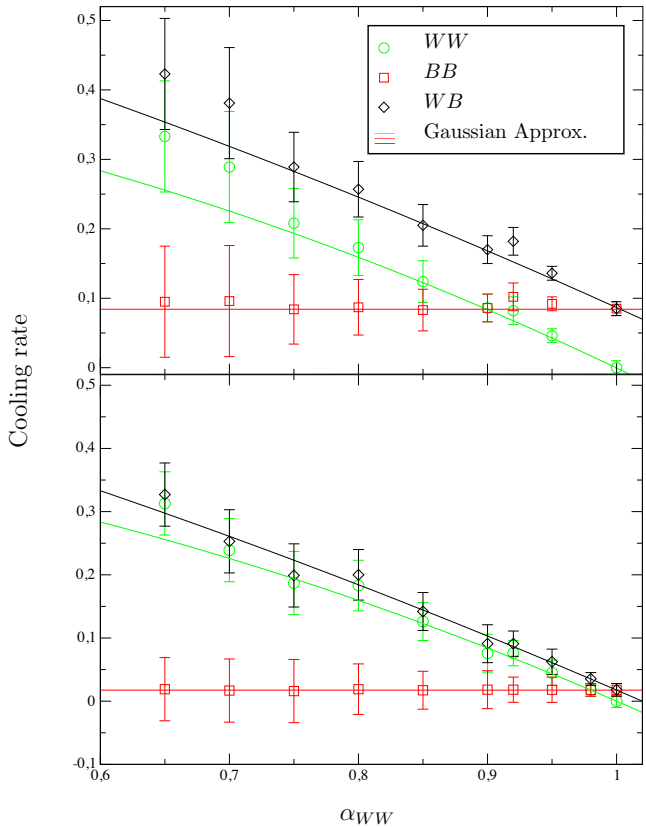}
 \caption{Values of the cooling rates considering the three types of collisions, $WW$, $BB$ and $WB$, for $\alpha_{BB}=0.9$ (top panel) and $\alpha_{BB}=0.98$ (bottom panel) and for $0.65 \le \alpha_{WW} \le 1.0$; the straight lines correspond to the Gaussian approximation for equal fractions of collisions. Curves for $BB$ collisions are flat since $\alpha_{BB}$ is kept fixed in the simulations.}
 \label{coolinww}
 \end{figure}

We further analyzed the cooling rates corresponding to the different types of collisions: single-color ($WW$ and $BB$) or two-color ($WB$).
These have been measured directly from the dissipated energy at given time intervals for $\alpha_{BB}=0.9$ and $\alpha_{BB}=0.98$ and varying $\alpha_{WW}$
between 0.65 and 1.0. The results are displayed in Fig.~\ref{coolinww}; note that the top panel  of this figure corresponds to the data reported in Fig.~\ref{coolin2}.

\begin{figure}[!t]
  \centering
 \includegraphics[angle=0,width=8.6cm]{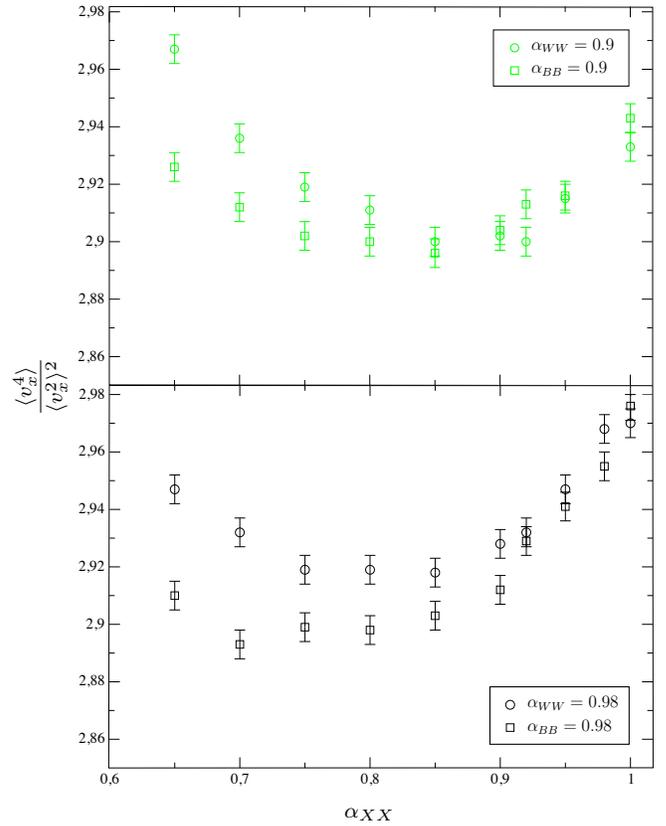}\\
  \caption{Values of the kurtosis for the collision types $WW$ and $BB$. The $X$ axis refers to varying values of either $\alpha_{WW}$ or $\alpha_{BB}$, when fixing the other. The top panel refers to the kurtosis when $\alpha_{WW}=0.9$ and  $\alpha_{BB}$ ranges from 0.6 to 1 (green circles) and when $\alpha_{BB}=0.9$ and  $\alpha_{WW}$ ranges from 0.6 to 1 (green squares). The same applies to the bottom panel, where each coefficient of restitution, when fixed, is 0.98. }
 \label{curtosis_w_b09}
 \end{figure}

 We have also measured the kurtosis of the different types of collisions for the same collection of values of the parameters $\alpha_{WW}$ and $\alpha_{BB}$. We observe that the results we obtain are also close to the Gaussian distribution, as shown in  Fig.~\ref{curtosis_w_b09}. Again, the results in Figs.~\ref{coolinww} and \ref{curtosis_w_b09} are found to be qualitatively similar to those reported in \cite{Brey96, Brey04}.

\begin{figure}[!h]
  \centering
  \includegraphics[angle=0,width=8.6cm]{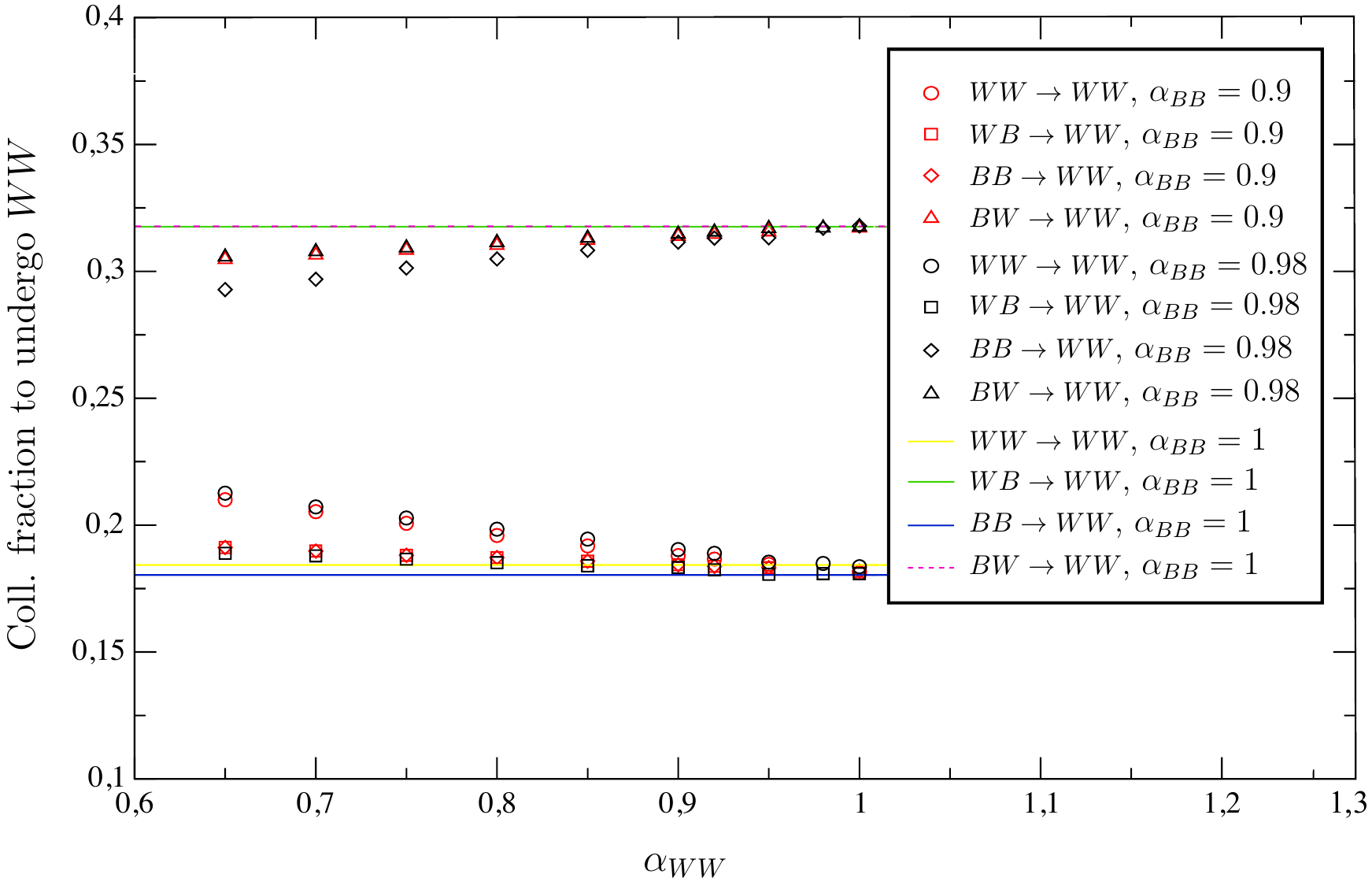}
 \caption{Values of the conditional probabilities for the particles that, having experienced a collision of a given type ($WW$, $WB$, $BB$ or $BW$) will later undergo a $WW$ collision. Here, the coefficients of restitutions are $\alpha_{BB}=0.9$, $0.98$ and $1$, whereas  $0.65 \le \alpha_{WW} \le 1.0$. The  lines refer to the elastic systems.}
 \label{corrrelacion}
 \end{figure}

Once the existence of a difference between the fraction of collisions of each type, the cooling rates and the temperatures has been noticed, the next natural step is to consider the precollisional correlations. The idea behind this analysis is to investigate whether the consideration of anisotropic disks may produce an effect on the dynamics of the collisions. We have measured the precollisional correlations for the three kinds of collision and taken as the reference system the one of colored but elastic disks. To that end, a numeric label, 1, 2, 3 or 4, is assigned to each disk after it collides, to account for the sides at contact. This label is checked against the value it had prior to collision, where we had made a distinction between the four possible types. We have found that there seems not to be a big effect as the values of the inelasticities are varied. In fact, there is only a small precollisional correlation of the colliding colors related to them, as shown in Fig.~\ref{corrrelacion}. In this figure, we have represented the conditional probabilities of having a $WW$ collision provided that the previous collision was of one of the four alternatives. Similar results are obtained for the other combinations, as reported in Tables I-III of the Appendix, where each entry in the ``C.Prob.'' columns provides conditional probability as the fraction of the corresponding collisions (for instance, when the collision is of the type $WW$ (1) and previously it has been of the type $WW$ (1), $WB$ (2), $BB$ (3) or $BW$ (4), and similarly for all suitable combinations).

 \section{Concluding Remarks} \label{s5}
 
In this paper we have addressed a system of inelastic Janus-like disks by means of extensive numerical experiments. In particular we have measured different statistical quantities, namely, temperature evolution, velocity distribution functions, kurtosis and precollisional correlations. We find that for this system a HCS also arises even in the presence of correlations. In particular, a Haff's law is obeyed by the temperature irrespective of the fact that the molecular chaos assumption does not hold here. As far as we are aware, this is the first time that such a system, along with some of its features, has been reported. This opens new avenues for research including, for instance, stationary states sustained by injection of energy, rough disks and  the analysis of problems like segregation \cite{GMV13,VG14}, memory effects \cite{PT14, TP14, LVPS17, LVPS19, TLLVPS19} and shear states \cite{G93,GS03}. Of particular interest are the studies of the clustering formation processes and the phase transitions that may arise as the density is increased. We plan to examine some of these problems in the near future.

 
\section*{Acknowledgements}

This paper is dedicated to the memory of Mar\'ia Jos\'e Ruiz Montero who together with J. J. Brey  originally suggested to one of us (A. L.) to undertake this project. This work has been partially supported by the Spanish Ministerio de Ciencia, Innovaci\'on y Universidades and the Agencia Estatal de Investigaci\'on Grants (partially financed by the ERDF) No. MTM2017-84446-C2-2-R (A.L. and A.T.) and No. FIS2017-84440-C2-2-P (A.T.). A. L.  also acknowledges the Program Jos\'e Castillejos grant No. CAS18/00335, which allowed him to visit the Instituto de Energ\'ias Renovables (U.N.A.M.) where this work was carried out.


\section*{Appendix}\label{appendix}

For completeness, we assemble in this appendix the conditional probabilities of a collision of type 1, 2, 3 or 4, provided that the preceding collision was of one of these four types, for different values of $\alpha_{BB}$ and $\alpha_{WW}$.
\begin{table}
 \caption{Precollisional correlations for $\alpha_{BB}$=0.9 and $0.98$ and for $\alpha_{WW}=0.65$, $0.7$ and $0.75$.\\}
\centering
\begin{tabular}{| c | c|c|c || c | c|c|c |}
\hline
$\alpha_{BB} $ & $ \alpha_{WW} $&\ Type \ &\  C.Prob. \ & $\alpha_{BB} $ & $ \alpha_{WW} $&\ Type \ &\  C.Prob. \ \\
\hline
\multirow{59}{*}{0.9}  & \multirow{20}{*}{0.65} & 1 $\rightarrow$ 1& 0.2100 & \multirow{59}{*}{0.98}  &\multirow{20}{*}{0.65}& 1 $\rightarrow$ 1& 0.2126\\
 & & 2 $\rightarrow$ 1& 0.1912 &   & & 2 $\rightarrow$ 1& 0.1889\\ 
 & & 3 $\rightarrow$ 1& 0.2941 &   & & 3 $\rightarrow$ 1& 0.2928\\
 & & 4 $\rightarrow$ 1& 0.3047 &   & & 4 $\rightarrow$ 1& 0.3057\\
 & & 1 $\rightarrow$ 2& 0.2043 &   & & 1 $\rightarrow$ 2& 0.2062\\
 & & 2 $\rightarrow$ 2& 0.1943 &   & & 2 $\rightarrow$ 2& 0.1921\\
 & & 3 $\rightarrow$ 2& 0.2957 &   & & 3 $\rightarrow$ 2& 0.2947\\
 & & 4 $\rightarrow$ 2& 0.3057 &   & & 4 $\rightarrow$ 2& 0.3070\\
 & & 1 $\rightarrow$ 3& 0.3194 &   & & 1 $\rightarrow$ 3& 0.3244\\
 & & 2 $\rightarrow$ 3& 0.3110 &   & & 2 $\rightarrow$ 3& 0.3136\\
 & & 3 $\rightarrow$ 3& 0.1800 &   & & 3 $\rightarrow$ 3& 0.1747\\
 & & 4 $\rightarrow$ 3& 0.1896 &   & & 4 $\rightarrow$ 3& 0.1872\\
 & & 1 $\rightarrow$ 4& 0.3190 &   & & 1 $\rightarrow$ 4& 0.3235\\
 & & 2 $\rightarrow$ 4& 0.3091 &   & & 2 $\rightarrow$ 4& 0.3118\\
 & & 3 $\rightarrow$ 4& 0.1774 &   & & 3 $\rightarrow$ 4& 0.1725\\
 & & 4 $\rightarrow$ 4& 0.1945 &   & & 4 $\rightarrow$ 4& 0.1921\\
\cline{2-4}\cline{6-8}
 &\multirow{20}{*}{0.7} & 1 $\rightarrow$ 1& 0.2053 & &\multirow{20}{*}{0.7}& 1 $\rightarrow$ 1& 0.2072\\
 & & 2 $\rightarrow$ 1& 0.1898 &   & & 2 $\rightarrow$ 1& 0.1879\\
 & & 3 $\rightarrow$ 1& 0.2984 &   & & 3 $\rightarrow$ 1& 0.2969\\
 & & 4 $\rightarrow$ 1& 0.3065 &   & & 4 $\rightarrow$ 1& 0.3078\\
 & & 1 $\rightarrow$ 2& 0.1999 &   & & 1 $\rightarrow$ 2& 0.2013\\
 & & 2 $\rightarrow$ 2& 0.1930 &   & & 2 $\rightarrow$ 2& 0.1908\\
 & & 3 $\rightarrow$ 2& 0.2999 &   & & 3 $\rightarrow$ 2& 0.2988\\
 & & 4 $\rightarrow$ 2& 0.3073 &   & & 4 $\rightarrow$ 2& 0.3090\\
 & & 1 $\rightarrow$ 3& 0.3179 &   & & 1 $\rightarrow$ 3& 0.3231\\
 & & 2 $\rightarrow$ 3& 0.3117 &   & & 2 $\rightarrow$ 3& 0.3144\\
 & & 3 $\rightarrow$ 3& 0.1817 &   & & 3 $\rightarrow$ 3& 0.1762\\
 & & 4 $\rightarrow$ 3& 0.1887 &   & & 4 $\rightarrow$ 3& 0.1861\\ 
 & & 1 $\rightarrow$ 4& 0.3177 &   & & 1 $\rightarrow$ 4& 0.3223\\
 & & 2 $\rightarrow$ 4& 0.3099 &   & & 2 $\rightarrow$ 4& 0.3127\\
 & & 3 $\rightarrow$ 4& 0.1791 &   & & 3 $\rightarrow$ 4& 0.1739\\
 & & 4 $\rightarrow$ 4& 0.1932 &   & & 4 $\rightarrow$ 4& 0.1909\\
\cline{2-4}\cline{6-8}
  &\multirow{20}{*}{0.75} & 1 $\rightarrow$ 1& 0.2007 &   &\multirow{20}{*}{0.75}& 1 $\rightarrow$ 1& 0.1971\\
  & & 2 $\rightarrow$ 1& 0.1881 &   & & 2 $\rightarrow$ 1& 0.1841\\
  & & 3 $\rightarrow$ 1& 0.3027 &   & & 3 $\rightarrow$ 1& 0.3031\\
  & & 4 $\rightarrow$ 1& 0.3084 &   & & 4 $\rightarrow$ 1& 0.3119\\
  & & 1 $\rightarrow$ 2& 0.1956 &   & & 1 $\rightarrow$ 2& 0.1921\\
  & & 2 $\rightarrow$ 2& 0.1916 &   & & 2 $\rightarrow$ 2& 0.1872\\
  & & 3 $\rightarrow$ 2& 0.3040 &   & & 3 $\rightarrow$ 2& 0.3045\\
  & & 4 $\rightarrow$ 2& 0.3088 &   & & 4 $\rightarrow$ 2& 0.3124\\
  & & 1 $\rightarrow$ 3& 0.3168 &   & & 1 $\rightarrow$ 3& 0.3184\\
  & & 2 $\rightarrow$ 3& 0.3122 &   & & 2 $\rightarrow$ 3& 0.3134\\
  & & 3 $\rightarrow$ 3& 0.1834 &   & & 3 $\rightarrow$ 3& 0.1785\\
  & & 4 $\rightarrow$ 3& 0.1875 &   & & 4 $\rightarrow$ 3& 0.1845\\
  & & 1 $\rightarrow$ 4& 0.3164 &   & & 1 $\rightarrow$ 4& 0.3207\\
  & & 2 $\rightarrow$ 4& 0.3111 &   & & 2 $\rightarrow$ 4& 0.3149\\
  & & 3 $\rightarrow$ 4& 0.1807 &   & & 3 $\rightarrow$ 4& 0.1771\\
  & & 4 $\rightarrow$ 4& 0.1919 &   & & 4 $\rightarrow$ 4& 0.1889\\
\hline
\end{tabular}
\end{table}


\begin{table}
 \caption{Precollisional correlations for $\alpha_{BB}$=0.9 and $0.98$ and for $\alpha_{WW}=0.8$, $0.85$ and $0.9$.\\}
\centering
\begin{tabular}{| c | c|c|c || c | c|c|c |}
\hline
$\alpha_{BB} $ & $ \alpha_{WW} $&\ Type \ &\  C.Prob. \ & $\alpha_{BB} $ & $ \alpha_{WW} $&\ Type \ &\  C.Prob. \ \\
\hline
\multirow{59}{*}{0.9}  & \multirow{20}{*}{0.8} & 1 $\rightarrow$ 1& 0.1959 & \multirow{59}{*}{0.98} & \multirow{20}{*}{0.8} & 1 $\rightarrow$ 1& 0.1984\\
  & & 2 $\rightarrow$ 1& 0.1872 &   & & 2 $\rightarrow$ 1& 0.1852\\
  & & 3 $\rightarrow$ 1& 0.3066 &   & & 3 $\rightarrow$ 1& 0.3049\\
  & & 4 $\rightarrow$ 1& 0.3103 &   & & 4 $\rightarrow$ 1& 0.3114\\
  & & 1 $\rightarrow$ 2& 0.1917 &   & & 1 $\rightarrow$ 2& 0.1933\\
  & & 2 $\rightarrow$ 2& 0.1904 &   & & 2 $\rightarrow$ 2& 0.1886\\
  & & 3 $\rightarrow$ 2& 0.3072 &   & & 3 $\rightarrow$ 2& 0.3063\\
  & & 4 $\rightarrow$ 2& 0.3107 &   & & 4 $\rightarrow$ 2& 0.3118\\
  & & 1 $\rightarrow$ 3& 0.3157 &   & & 1 $\rightarrow$ 3& 0.3205\\
  & & 2 $\rightarrow$ 3& 0.3129 &   & & 2 $\rightarrow$ 3& 0.3154\\
  & & 3 $\rightarrow$ 3& 0.1850 &   & & 3 $\rightarrow$ 3& 0.1798\\
  & & 4 $\rightarrow$ 3& 0.1863 &   & & 4 $\rightarrow$ 3& 0.1843\\
  & & 1 $\rightarrow$ 4& 0.3156 &   & & 1 $\rightarrow$ 4& 0.3202\\
  & & 2 $\rightarrow$ 4& 0.3120 &   & & 2 $\rightarrow$ 4& 0.3144\\
  & & 3 $\rightarrow$ 4& 0.1818 &   & & 3 $\rightarrow$ 4& 0.1770\\
  & & 4 $\rightarrow$ 4& 0.1907 &   & & 4 $\rightarrow$ 4& 0.1885\\
\hline
  & \multirow{20}{*}{0.85}& 1 $\rightarrow$ 1& 0.1918 &   &\multirow{20}{*}{0.85}& 2 $\rightarrow$ 1& 0.1840\\
  & & 2 $\rightarrow$ 1& 0.1858 &  & & 2 $\rightarrow$ 1& 0.1840\\
  & & 3 $\rightarrow$ 1& 0.3101 &   & & 3 $\rightarrow$ 1& 0.3083\\
  & & 4 $\rightarrow$ 1& 0.3122 &   & & 4 $\rightarrow$ 1& 0.3131\\
  & & 1 $\rightarrow$ 2& 0.1876 &   & & 1 $\rightarrow$ 2& 0.1898\\
  & & 2 $\rightarrow$ 2& 0.1893 &   & & 2 $\rightarrow$ 2& 0.1875\\
  & & 3 $\rightarrow$ 2& 0.3106 &   & & 3 $\rightarrow$ 2& 0.3094\\
  & & 4 $\rightarrow$ 2& 0.3125 &   & & 4 $\rightarrow$ 2& 0.3132\\
  & & 1 $\rightarrow$ 3& 0.3146 &   & & 1 $\rightarrow$ 3& 0.3195\\
  & & 2 $\rightarrow$ 3& 0.3136 &   & & 2 $\rightarrow$ 3& 0.3160\\
  & & 3 $\rightarrow$ 3& 0.1866 &   & & 3 $\rightarrow$ 3& 0.1810\\
  & & 4 $\rightarrow$ 3& 0.1851 &   & & 4 $\rightarrow$ 3& 0.1834\\
  & & 1 $\rightarrow$ 4& 0.3149 &   & & 1 $\rightarrow$ 4& 0.3190\\
  & & 2 $\rightarrow$ 4& 0.3126 &   & & 2 $\rightarrow$ 4& 0.3153\\
  & & 3 $\rightarrow$ 4& 0.1832 &   & & 3 $\rightarrow$ 4& 0.1778\\
  & & 4 $\rightarrow$ 4& 0.1894 &   & & 4 $\rightarrow$ 4& 0.1878\\
\hline
  & \multirow{20}{*}{0.9} & 1 $\rightarrow$ 1& 0.1880 &   & \multirow{20}{*}{0.9} & 1 $\rightarrow$ 1& 0.1904\\
  & & 2 $\rightarrow$ 1& 0.1844 &  & & 2 $\rightarrow$ 1& 0.1833\\
  & & 3 $\rightarrow$ 1& 0.3136 &   & & 3 $\rightarrow$ 1& 0.3115\\
  & & 4 $\rightarrow$ 1& 0.3139 &   & & 4 $\rightarrow$ 1& 0.3147\\
  & & 1 $\rightarrow$ 2& 0.1844 &   & & 1 $\rightarrow$ 2& 0.1862\\
  & & 2 $\rightarrow$ 2& 0.1882 &   & & 2 $\rightarrow$ 2& 0.1863\\
  & & 3 $\rightarrow$ 2& 0.3137 &   & & 3 $\rightarrow$ 2& 0.3125\\
  & & 4 $\rightarrow$ 2& 0.3136 &   & & 4 $\rightarrow$ 2& 0.3149\\
  & & 1 $\rightarrow$ 3& 0.3136 &   & & 1 $\rightarrow$ 3& 0.3182\\
  & & 2 $\rightarrow$ 3& 0.3140 &   & & 2 $\rightarrow$ 3& 0.3168\\
  & & 3 $\rightarrow$ 3& 0.1879 &   & & 3 $\rightarrow$ 3& 0.1824\\
  & & 4 $\rightarrow$ 3& 0.1846 &   & & 4 $\rightarrow$ 3& 0.1826\\
  & & 1 $\rightarrow$ 4& 0.3140 &   & & 1 $\rightarrow$ 4& 0.3185\\
  & & 2 $\rightarrow$ 4& 0.3135 &   & & 2 $\rightarrow$ 4& 0.3156\\
  & & 3 $\rightarrow$ 4& 0.1845 &   & & 3 $\rightarrow$ 4& 0.1792\\
  & & 4 $\rightarrow$ 4& 0.1880 &   & & 4 $\rightarrow$ 4& 0.1866\\
\hline
\end{tabular}
\end{table}


\begin{table}
 \caption{Precollisional correlations for $\alpha_{BB}$=0.9 and $0.98$ and for $\alpha_{WW}=0.92$, $0.95$ and $1$.\\}
\centering
\begin{tabular}{| c | c|c|c || c | c|c|c |}
\hline
$\alpha_{BB} $ & $ \alpha_{WW} $&\ Type \ &\  C.Prob. \ & $\alpha_{BB} $ & $ \alpha_{WW} $&\ Type \ &\  C.Prob. \ \\
\hline
\multirow{59}{*}{0.9}   &\multirow{20}{*}{0.92}& 1 $\rightarrow$ 1& 0.1866 & \multirow{59}{*}{0.98}   &\multirow{20}{*}{0.92}& 1 $\rightarrow$ 1& 0.1889\\
  & & 2 $\rightarrow$ 1& 0.1838 &   & & 2 $\rightarrow$ 1& 0.1824\\
  & & 3 $\rightarrow$ 1& 0.3149 &   & & 3 $\rightarrow$ 1& 0.3131\\
  & & 4 $\rightarrow$ 1& 0.3146 &   & & 4 $\rightarrow$ 1& 0.3155\\
  & & 1 $\rightarrow$ 2& 0.1829 &   & & 1 $\rightarrow$ 2& 0.1847\\
  & & 2 $\rightarrow$ 2& 0.1881 &   & & 2 $\rightarrow$ 2& 0.1862\\
  & & 3 $\rightarrow$ 2& 0.3147 &   & & 3 $\rightarrow$ 2& 0.3135\\
  & & 4 $\rightarrow$ 2& 0.3142 &   & & 4 $\rightarrow$ 2& 0.3155\\
  & & 1 $\rightarrow$ 3& 0.3130 &   & & 1 $\rightarrow$ 3& 0.3177\\
  & & 2 $\rightarrow$ 3& 0.3141 &   & & 2 $\rightarrow$ 3& 0.3169\\
  & & 3 $\rightarrow$ 3& 0.1888 &   & & 3 $\rightarrow$ 3& 0.1832\\
  & & 4 $\rightarrow$ 3& 0.1840 &   & & 4 $\rightarrow$ 3& 0.1821\\
  & & 1 $\rightarrow$ 4& 0.3134 &   & & 1 $\rightarrow$ 4& 0.3179\\
  & & 2 $\rightarrow$ 4& 0.3138 &   & & 2 $\rightarrow$ 4& 0.3162\\
  & & 3 $\rightarrow$ 4& 0.1848 &   & & 3 $\rightarrow$ 4& 0.1797\\
  & & 4 $\rightarrow$ 4& 0.1879 &   & & 4 $\rightarrow$ 4& 0.1862\\
\cline{2-4}\cline{6-8}
  &\multirow{20}{*}{0.95}& 1 $\rightarrow$ 1& 0.1845 &   &\multirow{20}{*}{0.95}& 1 $\rightarrow$ 1& 0.1854\\
  & & 2 $\rightarrow$ 1& 0.1832 &   & & 2 $\rightarrow$ 1& 0.1806\\
  & & 3 $\rightarrow$ 1& 0.3166 &   & & 3 $\rightarrow$ 1& 0.3132\\
  & & 4 $\rightarrow$ 1& 0.3156 &   & & 4 $\rightarrow$ 1& 0.3169\\
  & & 1 $\rightarrow$ 2& 0.1811 &   & & 1 $\rightarrow$ 2& 0.1817\\
  & & 2 $\rightarrow$ 2& 0.1870 &   & & 2 $\rightarrow$ 2& 0.1842\\
  & & 3 $\rightarrow$ 2& 0.3170 &   & & 3 $\rightarrow$ 2& 0.3136\\
  & & 4 $\rightarrow$ 2& 0.3148 &   & & 4 $\rightarrow$ 2& 0.3167\\
  & & 1 $\rightarrow$ 3& 0.3124 &   & & 1 $\rightarrow$ 3& 0.3154\\
  & & 2 $\rightarrow$ 3& 0.3145 &   & & 2 $\rightarrow$ 3& 0.3150\\
  & & 3 $\rightarrow$ 3& 0.1896 &   & & 3 $\rightarrow$ 3& 0.1827\\
  & & 4 $\rightarrow$ 3& 0.1834 &   & & 4 $\rightarrow$ 3& 0.1819\\
  & & 1 $\rightarrow$ 4& 0.3129 &   & & 1 $\rightarrow$ 4& 0.3181\\
  & & 2 $\rightarrow$ 4& 0.3141 &   & & 2 $\rightarrow$ 4& 0.3171\\
  & & 3 $\rightarrow$ 4& 0.1859 &   & & 3 $\rightarrow$ 4& 0.1807\\
  & & 4 $\rightarrow$ 4& 0.1870 &   & & 4 $\rightarrow$ 4& 0.1858\\  
\cline{2-4}\cline{6-8}
  &\multirow{20}{*}{1}& 1 $\rightarrow$ 1& 0.1812 &   &\multirow{20}{*}{1}& 1 $\rightarrow$ 1& 0.1836\\
  & & 2 $\rightarrow$ 1& 0.1819 &   & & 2 $\rightarrow$ 1& 0.1809\\
  & & 3 $\rightarrow$ 1& 0.3196 &   & & 3 $\rightarrow$ 1& 0.3179\\
  & & 4 $\rightarrow$ 1& 0.3171 &   & & 4 $\rightarrow$ 1& 0.3175\\
  & & 1 $\rightarrow$ 2& 0.1781 &   & & 1 $\rightarrow$ 2& 0.1799\\
  & & 2 $\rightarrow$ 2& 0.1862 &   & & 2 $\rightarrow$ 2& 0.1845\\
  & & 3 $\rightarrow$ 2& 0.3194 &   & & 3 $\rightarrow$ 2& 0.3185\\
  & & 4 $\rightarrow$ 2& 0.3162 &   & & 4 $\rightarrow$ 2& 0.3171\\
  & & 1 $\rightarrow$ 3& 0.3117 &   & & 1 $\rightarrow$ 3& 0.3169\\
  & & 2 $\rightarrow$ 3& 0.3149 &   & & 2 $\rightarrow$ 3& 0.3170\\
  & & 3 $\rightarrow$ 3& 0.1909 &   & & 3 $\rightarrow$ 3& 0.1855\\
  & & 4 $\rightarrow$ 3& 0.1825 &   & & 4 $\rightarrow$ 3& 0.1806\\
  & & 1 $\rightarrow$ 4& 0.3122 &   & & 1 $\rightarrow$ 4& 0.3172\\
  & & 2 $\rightarrow$ 4& 0.3150 &   & & 2 $\rightarrow$ 4& 0.3167\\
  & & 3 $\rightarrow$ 4& 0.1866 &   & & 3 $\rightarrow$ 4& 0.1814\\
  & & 4 $\rightarrow$ 4& 0.1861 &   & & 4 $\rightarrow$ 4& 0.1846\\
\hline
\end{tabular}
\end{table}

//






\bibliographystyle{apsrev}

\end{document}